\documentclass[twocolumn,twoside,slac_two]{revtex4}
\usepackage{graphicx}
\usepackage{fancyhdr}
\usepackage{amssymb}
\usepackage{amsbsy}
\begin{document}

\title{Simulation of relativistic shocks and associated radiation}

%

%

\author{K.-I. Nishikawa$^1$, J. Niemiec$^2$, M. Medvedev$^3$, B. Zhang$^4$, P. Hardee$^5$, Y. Mizuno$^1$,
 A. Nordlund$^6$, J. Frederiksen$^6$, H. Sol$^7$, M. Pohl$^8$, D. H. Hartmann$^{9}$,   J. F. Fishman$^{10}$}   
\affiliation{$^1$National Space Science and Technology Center,
  Huntsville, AL 35805, USA}
\affiliation{$^2$Institute of Nuclear Physics PAN, ul. Radzikowskiego 152, 31-342 Krak\'{o}w, Poland}
\affiliation{$^3$Department of Physics and Astronomy, University of Kansas, KS
66045, USA}
\affiliation{$^4$Department of Physics, University of Nevada, Las
Vegas, NV 89154, USA}
\affiliation{$^5$Department of Physics and Astronomy,
  The University of Alabama, Tuscaloosa, AL 35487, USA}
\affiliation{$^6$Niels Bohr Institute, University of Copenhagen, 
Juliane Maries Vej 30, 2100 Copenhagen \O, Denmark}
\affiliation{$^7$LUTH, Observatore de Paris-Meudon, 5 place Jules Jansen, 92195 Meudon Cedex, France}
\affiliation{$^8$Institue of Physics and Astronomy, University of Potsdam, Karl-Liebknecht-Strasse 24/25
14476 Potsdam-Golm
Germany}
\affiliation{$^9$Department of Physics and Astronomy, Clemson University, Clemson, SC 29634, USA}
\affiliation{$^{10}$NASA/MSFC,
  Huntsville, AL 35805, USA}   

\begin{abstract}
Using our new 3-D relativistic electromagnetic particle (REMP) code
parallelized with MPI, we investigated long-term particle
acceleration associated with a relativistic electron-positron jet
propagating in an unmagnetized ambient electron-positron plasma. 
We have also performed simulations with electron-ion jets. The
simulations were performed using a much longer simulation system
than our previous simulations in order to investigate the full
nonlinear stage of the Weibel instability for electron-positron jets and its particle
acceleration mechanism. Cold jet electrons are thermalized and ambient
electrons are accelerated in the resulting shocks for both cases. Acceleration of
ambient electrons leads to a maximum ambient electron density three
times larger than the original value for pair plasmas. Behind the bow shock in the jet
shock strong electromagnetic fields are generated.  These fields may
lead to time dependent afterglow emission. We calculated radiation from 
electrons propagating in a uniform parallel magnetic field to verify the 
technique. We also used the new technique to calculate emission from 
electrons based on simulations with a small system with two different cases for 
Lorentz factors (15 and 100). We obtained spectra 
which are consistent with those generated from electrons propagating
in turbulent magnetic fields with red noise. This turbulent magnetic field 
is similar to the magnetic field generated at an early nonlinear stage
of the Weibel instability. 
\end{abstract}

\maketitle

\thispagestyle{fancy}


\section{RPIC SIMULATIONS}
Particle-in-cell (PIC) simulations can shed light on the physical
mechanism of particle acceleration that occurs in the complicated
dynamics within relativistic shocks.  Recent PIC simulations of
relativistic electron-ion and electron-positron jets injected into an
ambient plasma show that acceleration occurs within the downstream jet
~\cite{nishi03,silva03,fred04,nishi05,hede05,jaro05,nishi06,ram07,chang08,anat08a,dsd08,anat08b,sironi09m}.
In general, these  simulations have confirmed that relativistic jets excite 
the Weibel instability, which generates current filaments and associated
magnetic fields ~\cite{weib59,medv99}, and accelerates electrons 
~\cite{hede05,nishi06,ram07,chang08,anat08a,anat08b,sironi09m}.

Therefore, the investigation of radiation resulting from accelerated particles 
(mainly electrons and positrons) in turbulent magnetic fields is essential for 
understanding radiation mechanisms and their observable spectral properties. 
In this report we present a new numerical method to obtain spectra from particles 
self-consistently traced in our PIC simulations.

\subsection{Relativistic  Jets
Injected into Unmagnetized Plasmas using a Large System}

We have performed simulations using a system with ($L_{\rm x}, L_{\rm y}, L_{\rm z}) 
= (4005\Delta,$ $131\Delta, 131\Delta)$ ($\Delta = 1$: grid size) and a total of
$\sim 1$ billion particles (12 particles $/$cell$/$species for the
ambient plasma) in the active grid zones ~\cite{nishi09b}. We have performed two
kind of simulations: an electron-positron jet is injected into an ambient pari plasma and an electron-ion
jet into electron-ion ambient plasma ($m_{\rm i}/m_{\rm e} = 20$).  In the
simulations the electron skin depth, $\lambda_{\rm ce} = c/\omega_{\rm
pe} = 10.0\Delta$, where $\omega_{\rm pe} = (4\pi e^{2}n_{\rm
e}/m_{\rm e})^{1/2}$ is the electron plasma frequency and the electron
Debye length $\lambda_{\rm e}$ is half of the grid size. Here the
computational domain is six times longer than in our previous
simulations ~\cite{nishi06,ram07}.  The electron number density of the
jet is $0.676n_{\rm e}$, where $n_{\rm e}$ is the ambient electron
density and $\gamma = 15$ for both cases.  The electron/positron thermal velocity of the
jet is $v^{\rm e}_{\rm j,th} = 0.014c$, the ion thermal velocity where $c = 1$ is the speed of light. 
The ion thermal velocity of jet is $v^{\rm i}_{\rm j,th} = v^{\rm e}_{\rm j,th}*\sqrt{m_{\rm e}/m_{\rm i}} = 
0.00313c$.
The electron/positron thermal velocity in the ambient plasma is 
$v^{\rm e}_{\rm a,th} = 0.05c$ ($v^{\rm i}_{\rm a,th} = 0.0112c$).

\begin{figure}[!h]
\centering
\includegraphics[width=80mm]{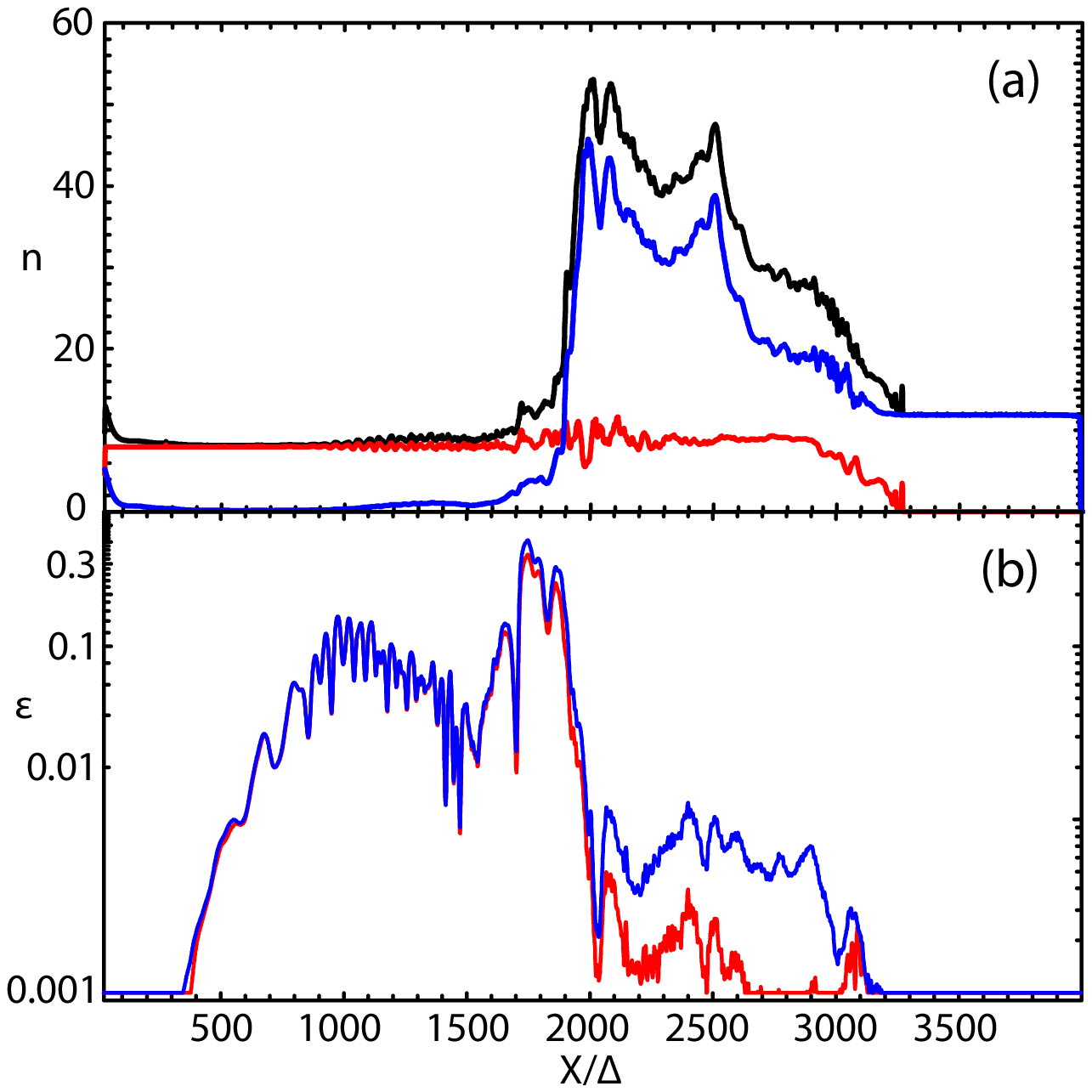}                            
\includegraphics[width=80mm]{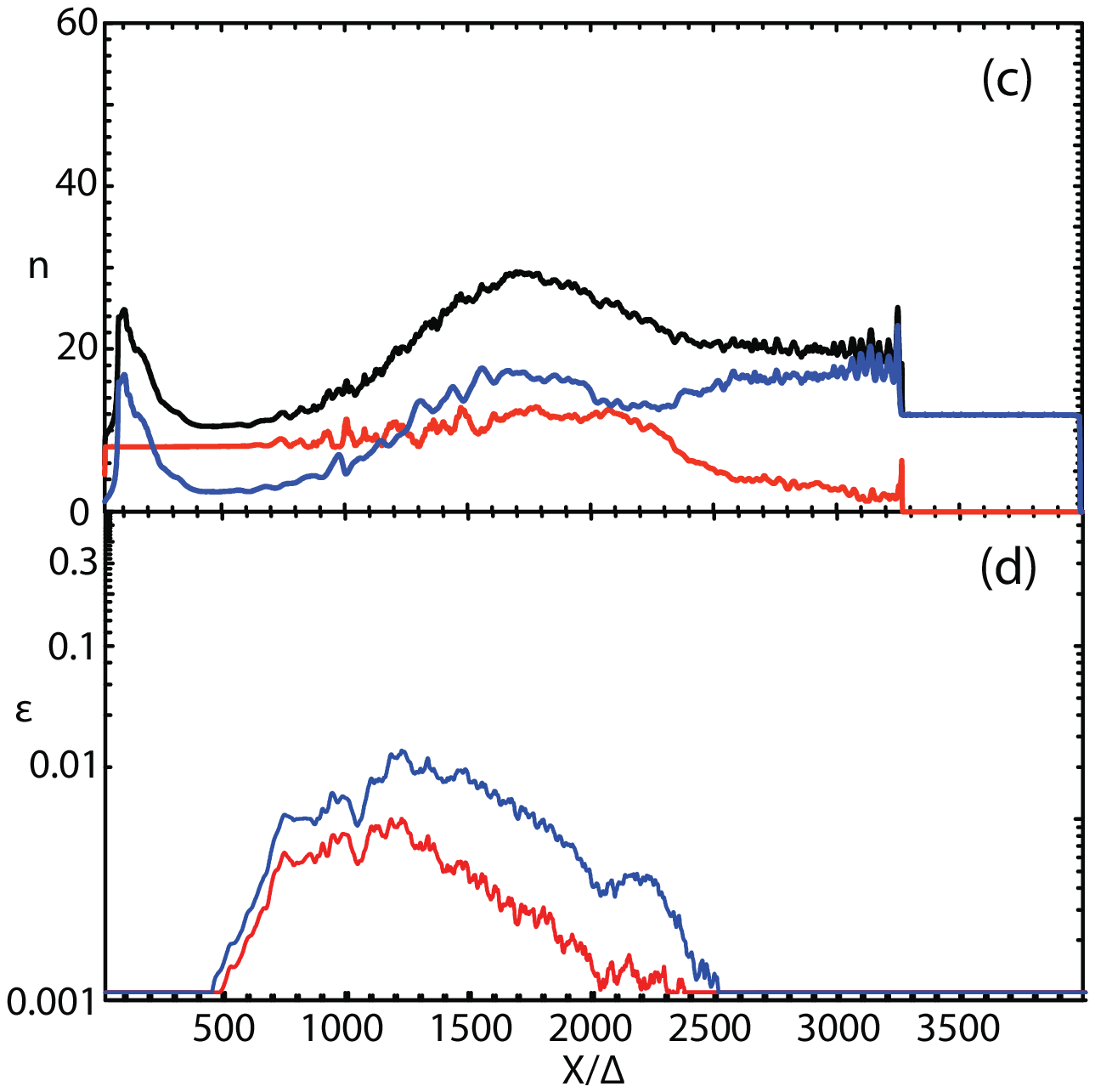}            
\vspace*{-.0cm}
\caption{The averaged values of electron density (a, c) and field energy
(b, d) along the $x$ at $t = 3750\omega_{\rm pe}^{-1}$. The left panel (Figs. 1a and 1b) is for the case
of pair plasmas and the right panel (Figs. 1c and 1d) is for the electron-ion case ($m_{\rm i}/m_{\rm e} =
20$). Figs. 1a and 1c
show jet electrons (red), ambient electrons (blue),
and the total electron density (black). 
Figs. 1b and 1d show electric field energy (red) and
magnetic field energy (blue) divided by the total kinetic energy.} 
\end{figure}

Figure 1 shows the averaged (in the $y-z$ plane) electron density and
electromagnetic field energy along the jet at  $3750\omega_{\rm pe}^{-1}$
for electron-positron (Figs. 1a and 1b) and electron-ion  (Figs. 1c and 1d) jets.  
The resulting profiles of jet (red), ambient (blue), and total (black) electron 
density are shown in Figs. 1a and 1c.  The ambient electrons are accelerated by the jet
electrons and pile up towards the front part of jet.  For electron-positron jet, at the earlier time
the ambient plasma density increases linearly behind the jet front. 
At the later time the ambient plasma
shows a rapid increase to a plateau behind the jet front, with
additional increase to a higher plateau farther behind the jet front.
The jet density remains approximately constant except near the jet front.

The acceleration of ambient electrons becomes visible when jet 
electrons pass about $x/\Delta = 500$. The maximum density of accelerated 
ambient electrons is attained at $t = 1750\omega_{\rm pe}^{-1}$.
The maximum density gradually reaches a plateau as seen in Fig. 1a. 
The maximum electromagnetic field energy is located at $x/\Delta = 1,700$ as shown in Fig. 1b.

The Weibel instability remains excited by continuously injected
jet particles and the electromagnetic fields are maintained at a high level,
about four times that seen in a previous, much shorter grid
simulation system ($L_{\rm x} = 640\Delta$). At the earlier simulation
time  a large electromagnetic structure is generated and accelerates the ambient
plasma. As shown in Fig. 1b, at the later simulation time the strong magnetic field 
extends up to $x/\Delta = 2,000$. These strong fields become very small beyond $x/\Delta = 
2000$ in the shocked ambient region
~\cite{nishi06,ram07}.

In the case with electron-ion jet, due to the heavier ions at $3750\omega_{\rm pe}^{-1}$
the density is piled up only slightly (Fig. 1c). The generated electromagnetic fields are smaller than those
for the electron-positron case. Furthermore, in the front of electron-ion jet the electromagnetic fields 
disappear as shown in Fig. 1d. In order to generate a shock it will require at least several time longer simulations.

\section{THE STANDARD SYNCHROTRON RADIATION MODEL}

A synchrotron shock model is widely adopted to describe the
radiation mechanism in the external shock thought to be responsible
for observed broad-band GRB afterglows ~\cite{zm04,p05a,p05b,zhangr07,nakar07}.
Associated
with this model are three major assumptions that are adopted in almost
all current GRB afterglow models. Firstly, electrons are assumed
to be ``Fermi'' accelerated at the relativistic shocks and to have a
power-law distribution with a power-law index $p$ upon acceleration,
i.e. $N(E_{\rm e})dE_{\rm e} \propto E^{-p} dE_{\rm e}$. 
This is consistent with recent PIC simulations of the shock formation and
particle acceleration ~\cite{anat08b} and also some Monte Carlo models 
~\cite{Achterberg01,ellison02,lemoi03}, but see 
~\cite{niemiec06,niemiecp06}.
Secondly, a fraction $\epsilon_{\rm e}$ (generally taken to be
$\leq 1$) of the total electrons associated with ISM baryons are
accelerated, and the total electron energy is a fraction $\epsilon_{\rm e}$ 
of the total internal energy in the shocked region. Thirdly, the
strength of the magnetic fields in the shocked region is unknown, but
its energy density ($B^{2}/8\pi$) is assumed to be a fraction
$\epsilon_{B}$ of the internal energy. These assumed ``micro-physics''
parameters, $p, \epsilon_{\rm e}$ and $\epsilon_{\rm B}$, whose values are
obtained from spectral fits ~\cite{paku01,yost03}
reflect a lack of knowledge of the underlying microphysics ~\cite{wax06}. 

The typical observed emission frequency from an electron with
(comoving) energy $\gamma_{\rm e}m_{\rm e}c^{2}$ in a frame with a
bulk Lorentz factor $\Gamma$ is $\nu = \Gamma\gamma_{\rm
e}^{2}(eB/2\pi m_{\rm e}c)$. Three critical frequencies are defined
by three characteristic electron energies. These are $\nu_{\rm m}$
(the injection frequency), $\nu_{\rm c}$ (the cooling frequency),
and $\nu_{\rm M}$ (the maximum synchrotron frequency). In our 
simulations of GRB afterglows, there is one additional relevant frequency, 
$\nu_{\rm a}$, due to synchrotron self-absorption at lower frequencies
~\cite{mrw98,spn98,zhangr07,nakar07}.

The general agreement between the blast wave dynamics and 
the direct measurements of the fireball size argue for
the validity of this model's dynamics ~\cite{zhangr07,nakar07}. 
The shock is most likely collisionless, i.e. mediated by plasma
instabilities ~\cite{wax06}. 
The electromagnetic instabilities
mediating the afterglow shock are expected to generate magnetic
fields. Afterglow radiation was therefore predicted to result from
synchrotron emission of shock accelerated electrons ~\cite{mr97}. 
The observed spectrum of afterglow radiation is indeed
remarkably consistent with synchrotron emission of electrons
accelerated to a power-law distribution, providing support for
the standard afterglow model based on synchrotron emission of shock 
accelerated electrons ~\cite{p99,p00,p05a,zm04,mes02,mes06,zhangr07,nakar07}.

In order to determine the luminosity and spectrum of synchrotron
radiation, the strength of the magnetic field ($\epsilon_{\rm B}$)
and the energy distribution of the electrons ($p$) must be
determined. Due to the lack of a first principles theory of
collisionless shocks, a purely phenomenological approach to the
model of afterglow radiation was ascribed without investigating in 
detail the processes responsible for particle acceleration and
magnetic field generation ~\cite{wax06}. 
Rather, one simply assumes
that a fraction $\epsilon_{\rm B}$ of the post-shock thermal energy
density is carried by the magnetic field, that a fraction
$\epsilon_{\rm e}$ is carried by electrons, and that the energy
distribution of the electrons is a power-law, $d \log n_{\rm e}/d
\log \varepsilon = p$ (above some minimum energy $\varepsilon_0$
which is determined by $\epsilon_{\rm e}$ and $p$), $\epsilon_{\rm
B}$, $\epsilon_{\rm e}$ and $p$ are treated as free parameters, 
determined by observations. It is important to clarify here that
the constraints implied on these parameters by the observations are
independent of any assumptions regarding the nature of the afterglow
shock and the processes responsible for particle acceleration or magnetic field 
generation. Any model should satisfy these observational constraints.

The properties of synchrotron (or ``jitter") 
emission from relativistic shocks will be determined by the magnetic field 
strength and structure and the electron energy distribution behind the shock.
The characteristics of jitter radiation may be important to
understanding the complex time evolution and/or spectral structure in
gamma-ray bursts ~\cite{pre98}. 
For example, jitter radiation
has been proposed as a means to explain GRB spectra below the peak frequency that are  
harder than the ``line of death'' spectral index associated with
synchrotron emission ~\cite{medv00,medv06},  i.e., 
the observed spectral power scales as $F_{\nu} \propto \nu^{2/3}$, whereas 
synchrotron spectra are $F_{\nu} \propto \nu^{1/3}$ or softer ~\cite{medv06}. 
Thus, it is essential to calculate radiation production by tracing electrons
(positrons) in self-consistently treated small-scale electromagnetic fields.

\section{NEW NUMERICAL METHOD FOR CALCULATING SYNCHROTRON EMISSION}

Let a particle be at position ${\bf{r}_{0}}(t)$ at
time $t$  ~\cite{nishi09a,hedeT05,hedeN05}. At the same time, 
we observe the electric field from the particle from position $\bf{r}$. However,
because of the finite velocity of light, we observe the
particle at an earlier position $\bf{r}_{0}(\rm{t}^{'})$ where it was at
the retarded time $t^{'} = t - \delta t^{'} = t - \bf{R}(\rm{t}^{'})/c$. Here
$\bf{R}(\rm{t}^{'}) = |\bf{r} - \bf{r}_{0}(\rm{t}^{'})|$ is the distance from the
charge (at the retarded time $t^{'}$) to the observer.

After some calculation and simplifying assumptions the total energy
$W$ radiated per unit solid angle per unit frequency from a charged
particle moving with instantaneous velocity $\boldsymbol{\beta}$ under
acceleration $\boldsymbol{\dot{\beta}}$ can be expressed as ~\cite{rybi79,jack99}

\begin{eqnarray}
&\frac{d^{2}W}{d\Omega d\omega} & =  \\
&\frac{\mu_{0} c
q^{2}}{16\pi^{3}} &\left|\int^{\infty}_{\infty}\frac{\bf{n}\times
[(\bf{n}-\mathbf{\beta})\times \dot{\bf{\beta}}]}{(1-\bf{\beta}\cdot
\bf{n})^{2}} 
 e^{i\omega(t^{'} -\bf{n} \cdot \bf{r}_{0}(t^{'})/c)}
dt^{'}\right|^{2} \nonumber
\end{eqnarray}
Here, $\bf{n} \equiv \bf{R}(\rm{t}^{'})/ |\bf{R}(\rm{t}^{'})|$ is a
unit vector that points from the particle's retarded position towards
the observer. 

The observer's viewing angle is set by the choice of 
$\bf{n}$ ($n_{\rm x}^{2}+n_{\rm y}^{2}+n_{\rm z}^{2} = 1$). 
The choice of unit vector $\bf{n}$ along the direction of propagation of
the jet (hereafter taken to be the $x$-axis) corresponds to head-on emission. 
For any other choice of $\bf{n}$ (e.g., $\theta_{\gamma} = 1/\gamma$), off-axis emission 
is seen by the observer. 

\subsection{Synchrotron radiation from two electrons propagating in parallel uniform magnetic field}

In order to calculate radiation from relativistic jets propagating along the $x$ 
direction ~\cite{nishi09a} we consider a test case which includes a parallel 
magnetic field ($B_{\rm x}$), and jet velocity of $v_{\rm j1,2} = 0.99c$. Two electrons 
are injected with different perpendicular velocities ($v_{\perp 1} = 0.1c, v_{\perp 2} 
= 0.12c$). A maximum Lorenz factor of 
$\gamma_{\max} =\{(1 - (v_{\rm j2}^{2} +v_{\perp 2}^{2})/c^{2}\}^{-1/2}  
= 13.48$  is calculated with the larger perpendicular velocity.

\begin{figure}[!h]
\centering
\includegraphics[width=60mm]{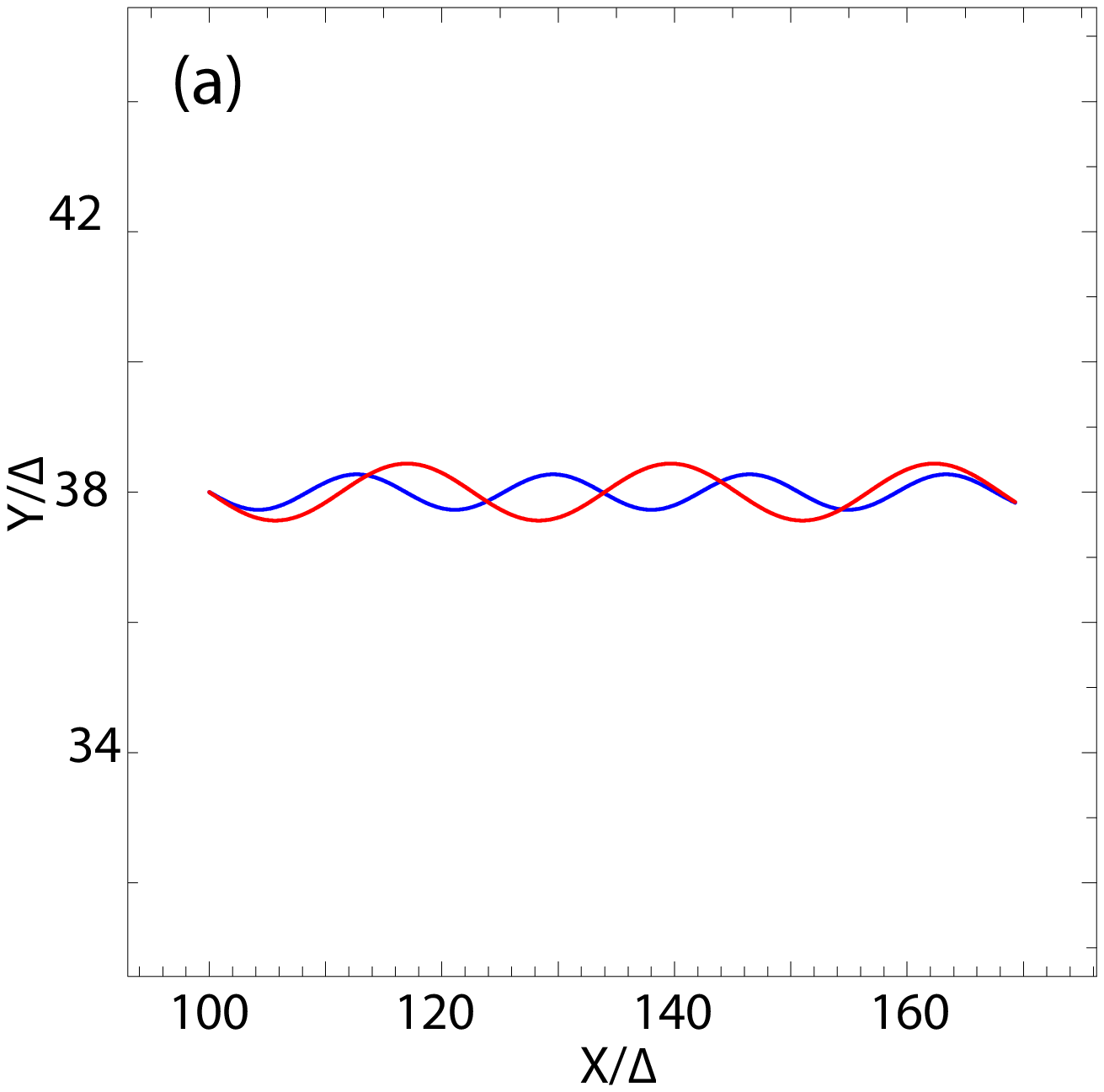}      
\includegraphics[width=65mm]{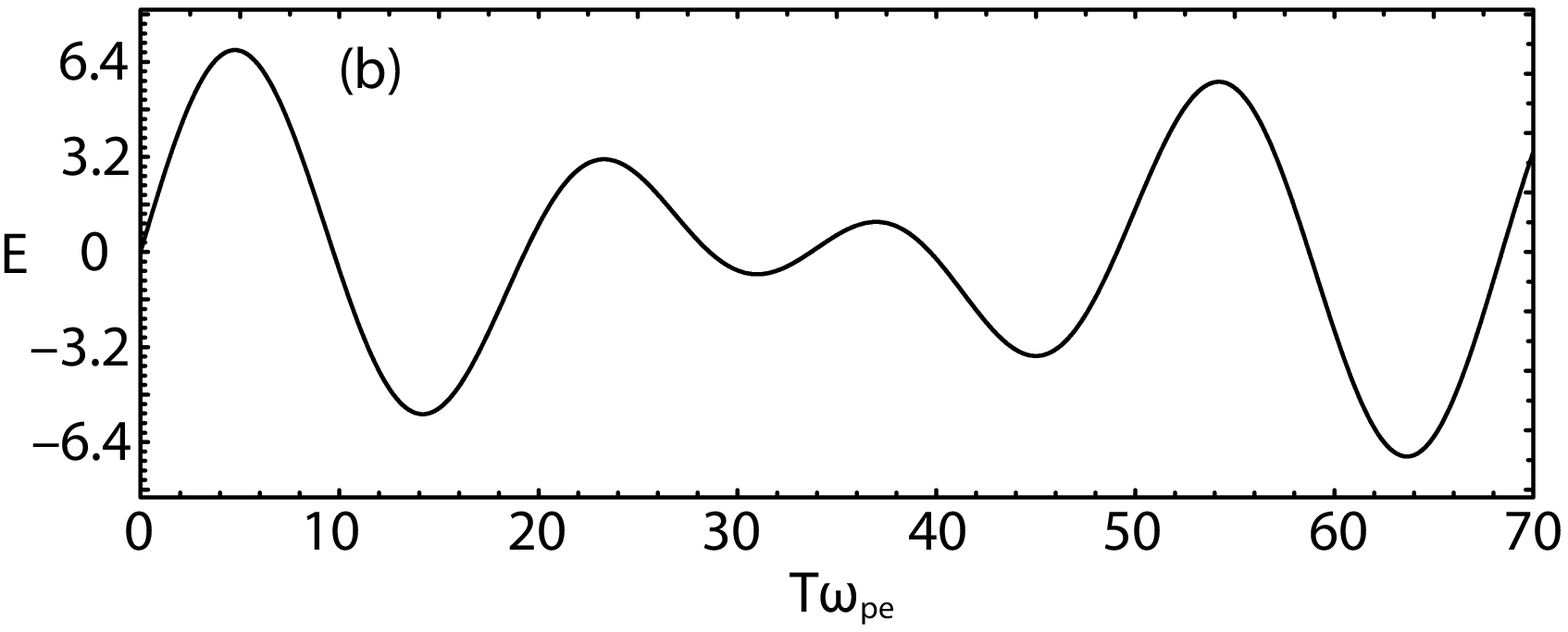}
\includegraphics[width=75mm]{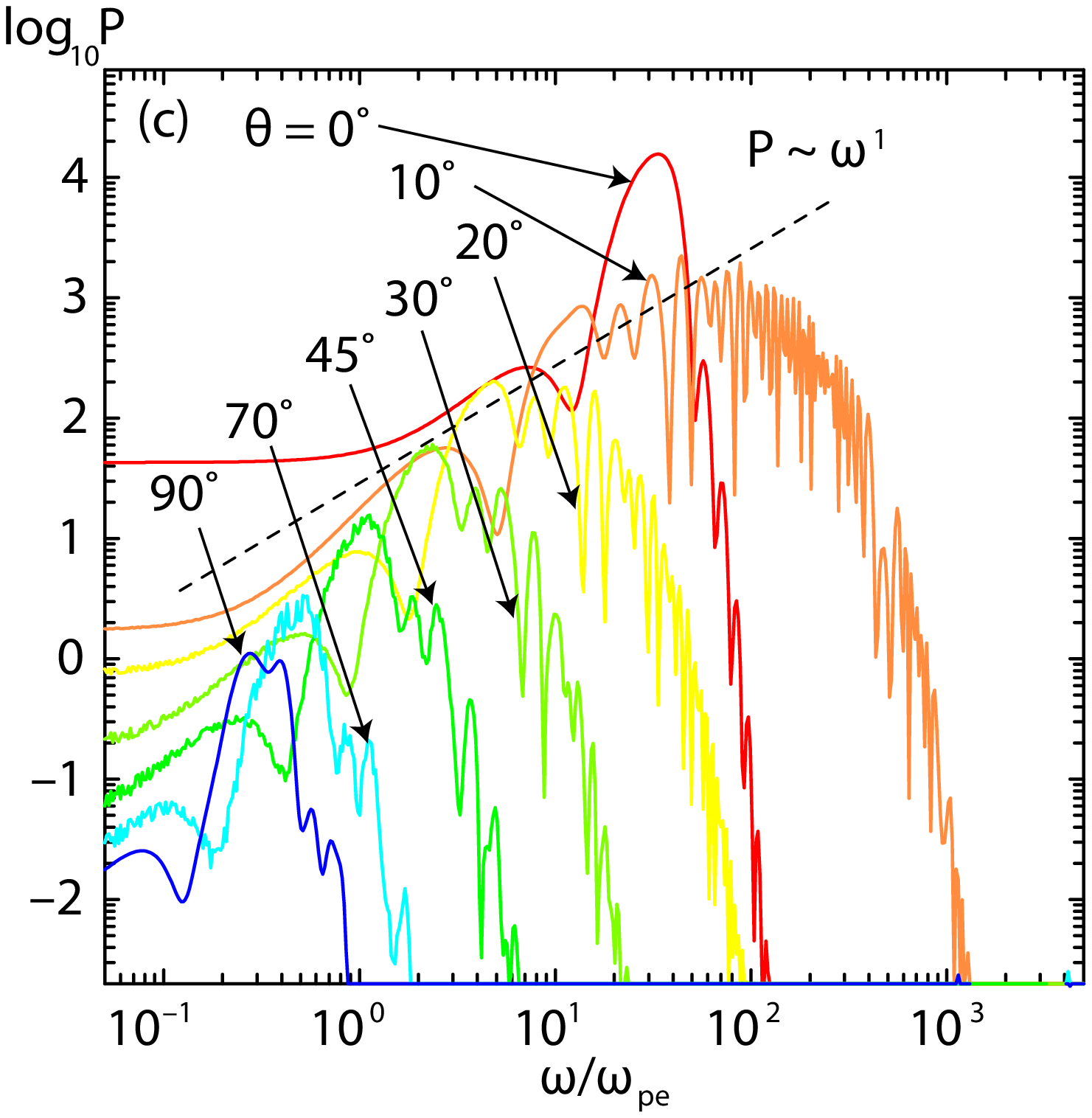}     
\vspace*{-0.cm}
\caption{The case with a strong magnetic field ($B_{\rm x} = 3.7$) and 
larger perpendicular velocity  ($v_{\perp 1} = 0.1c, v_{\perp 2}  = 0.12c$). The paths of 
two electrons moving helically along the $x-$direction in a homogenous magnetic 
field shown in the $x-y$-plane (a). The two electrons radiate a time dependent electric field. 
An observer situated at great distance along the n-vector sees the retarded electric field from
the moving electrons at the rest frame (b). 
The observed power spectrum at different viewing angles from the two electrons (c). 
Frequency is in units of $\omega_{\rm pe}^{-1}$. }
\end{figure}

Figure 2 shows electron trajectories in the $x - y$ plane 
(red: $v_{\perp 2} = 0.12c$, blue: $v_{\perp 1} = 0.1c$) (a: left
panel), the radiation (retarded) electric field  (b: middle panel), and spectra
(right panel) for the case $B_{\rm x} = 3.70$. The two electrons are
propagating left to right with gyration in the $y - z$ plane (not
shown). The gyroradius is about $0.44\Delta$ for the electron with the
larger perpendicular velocity. 
The seven curves show the power spectrum at viewing
angles of 0$^{\circ}$ (red), 10$^{\circ}$ (orange), 20$^{\circ}$ (yellow),
30$^{\circ}$ (moss green), 45$^{\circ}$ (green), 70$^{\circ}$ (light
blue), and 90$^{\circ}$ (blue). The higher frequencies become stronger
at the $10^{\circ}$ viewing angle. The critical angle for off-axis
radiation $\theta_{\gamma} = 180^{\circ}/(\pi \gamma_{\max})$ for this case is
4.25$^{\circ}$. As shown in this panel, the spectrum at a larger
viewing angle ($> 20^{\circ}$) has smaller amplitude.

Since the jet plasma has a large velocity $x$-component in the
simulation frame, the radiation from the particles
(electrons and positrons) is strongly beamed along the $x$-axis
(jitter radiation) ~\cite{medv00,medv06}. 

Equations 6.30a and 6.30b show that the radiation with the viewing angle $\alpha = 0$
disappears (see Fig. 6.5 in the textbook of Rybicki and Lightman ~\cite{rybi79}). 
However, based on other textbooks, radiation at the viewing angle $0^{\circ}$ should 
not vanish ~\cite{jack99,beke66,land80}. This aspect is shown in Fig. 2c, and at the higher frequency 
the amplitude at the viewing angle $10^{\circ}$ is stronger than that with viewing angle $0^{\circ}$.

\subsection{Calculating Synchrotron
and Jitter Emission from Electron Trajectories in Self-consistently
Generated Magnetic Field}

In order to validate our numerical method we performed simulations using 
a small system with ($L_{\rm x}, L_{\rm
y}, L_{\rm z}) = (645\Delta, 131\Delta, 131\Delta)$ ($\Delta = 1$: grid size) 
and a total of $\sim 0.5$ billion particles (12 particles$/$cell$/$species for the
ambient plasma) in the active grid zones ~\cite{nishi06}. First we performed simulations 
without calculating radiation up to $t = 450\omega_{\rm pe}^{-1}$. The jet front is 
located around about $x/\Delta = 480$. We selected 12,150  electrons for each jet and  
ambient electrons randomly.  Recently, a similar calculation has been carried out for the radiation from 
accelerated electrons in laser-wakefield acceleration ~\cite{martins09} and in shocks ~\cite{sironi09j}.

Figure 3 shows (a) the current filaments generated by the Weibel instability and (b) the 
phase space of $x/\Delta - \gamma V_{\rm x}$ for jet electrons (red) and ambient electrons 
(blue) at  $t = 450\omega_{\rm pe}^{-1}$ for $\gamma = 15$.

\begin{figure}[!h]
\centering
\includegraphics[width=80mm]{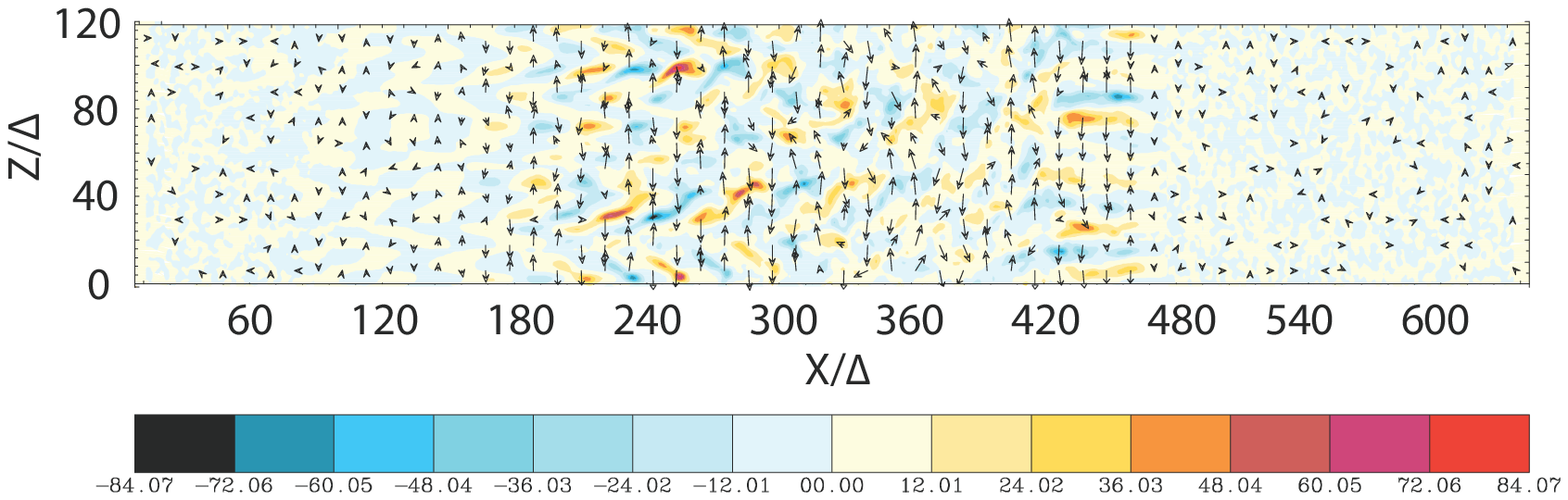}
\includegraphics[width=80mm]{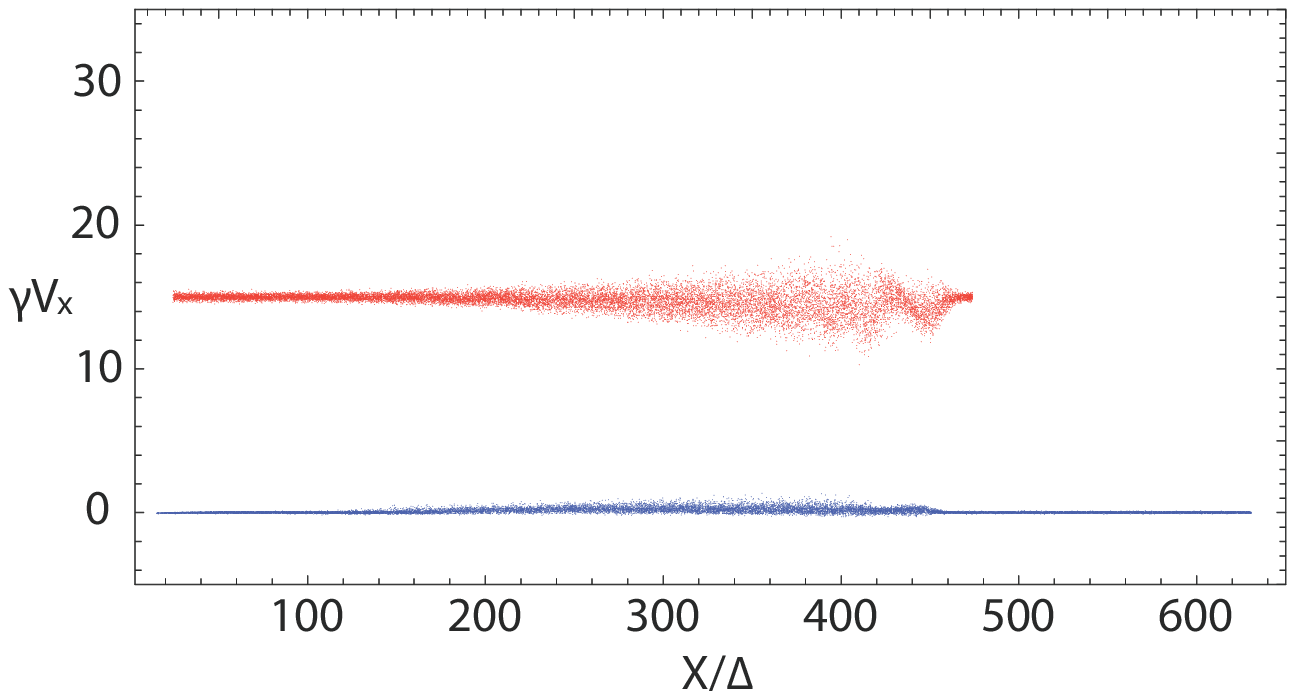}
\vspace{-0.5cm}
\caption{Two-dimensional images in the $x-z$ plane at $y/\Delta = 65$ for $t =450 \omega_{\rm pe}^{-1}$
for the case with $\gamma = 15$. 
The colors indicate the x-component of current density generated by the Weibel instability, with
the x- and z-components of magnetic field represented by arrows (a). Phase space distributions 
as a function of $x/\Delta-\gamma v_{\rm x}$ plotted for the jet (red) and ambient  (blue)
electrons at the same time.}
\end{figure}

\begin{figure}[!h]
\centering
\includegraphics[width=80mm]{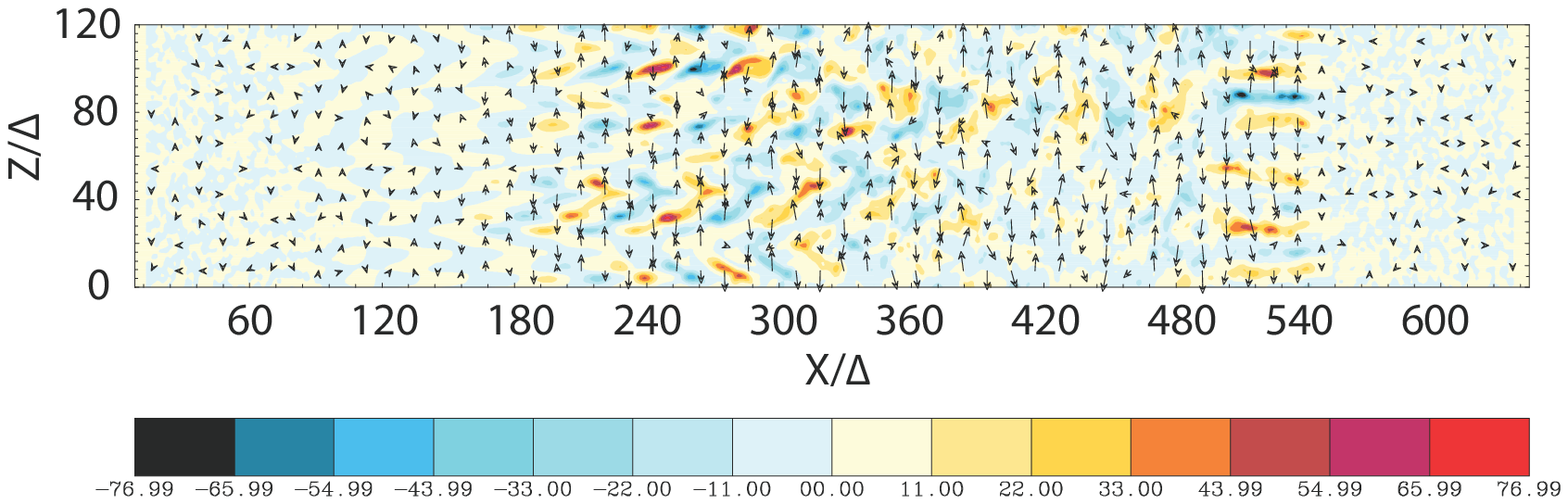}
\includegraphics[width=80mm]{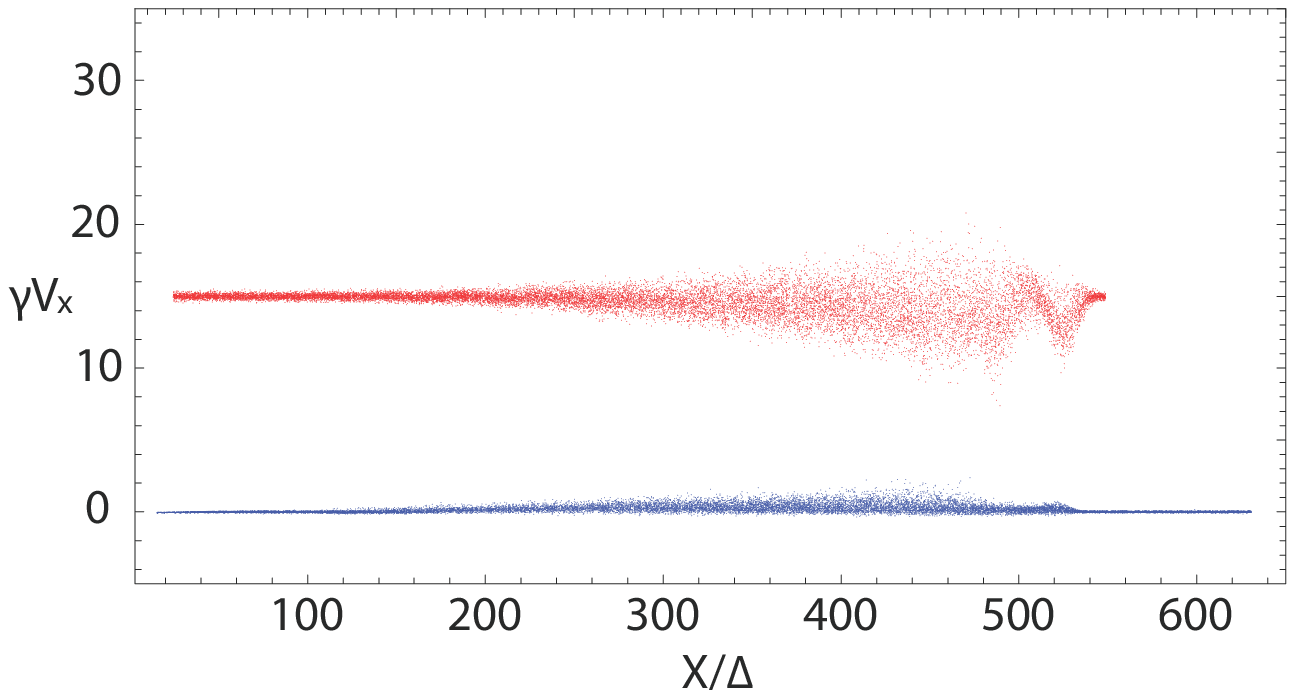}
\vspace{-0.cm}
\caption{Two-dimensional images in the $x-z$ plane at $y/\Delta = 65$ for $t =525 \omega_{\rm pe}^{-1}$ 
for the case with $\gamma = 15$.  The colors indicate
the x-component of current density generated by the Weibel instability, with
the x- and z-components of magnetic field represented by arrows (a).
Phase space distributions as a function of $x/\Delta-\gamma v_{\rm x}$ plotted for the jet (red) and 
ambient  (blue) electrons at the same time.}
\end{figure}

Figure 4 shows (a) the $x$-component of current density generated by the Weibel instability  
and (b) the phase space of jet electrons and ambient electrons after 
$t_{\rm s} = 75\omega_{\rm pe}^{-1}$ (at $t = 525\omega_{\rm pe}^{-1}$).

We calculated the emission from 12,150 electrons during the sampling time $t_{\rm s} = t_{\rm 2} -
t_{\rm 1} = 75\omega_{\rm pe}^{-1}$ with Nyquist frequency 
$\omega_{\rm N} = 1/2\Delta t = 200\omega_{\rm pe}$ where
$\Delta t = 0.005\omega_{\rm pe}^{-1} $ is the simulation time step and the frequency resolution
$\Delta \omega = 1/t_{\rm s} = 0.0133\omega_{\rm pe}$. 

\begin{figure}[!h]
\centering
\includegraphics[width=80mm]{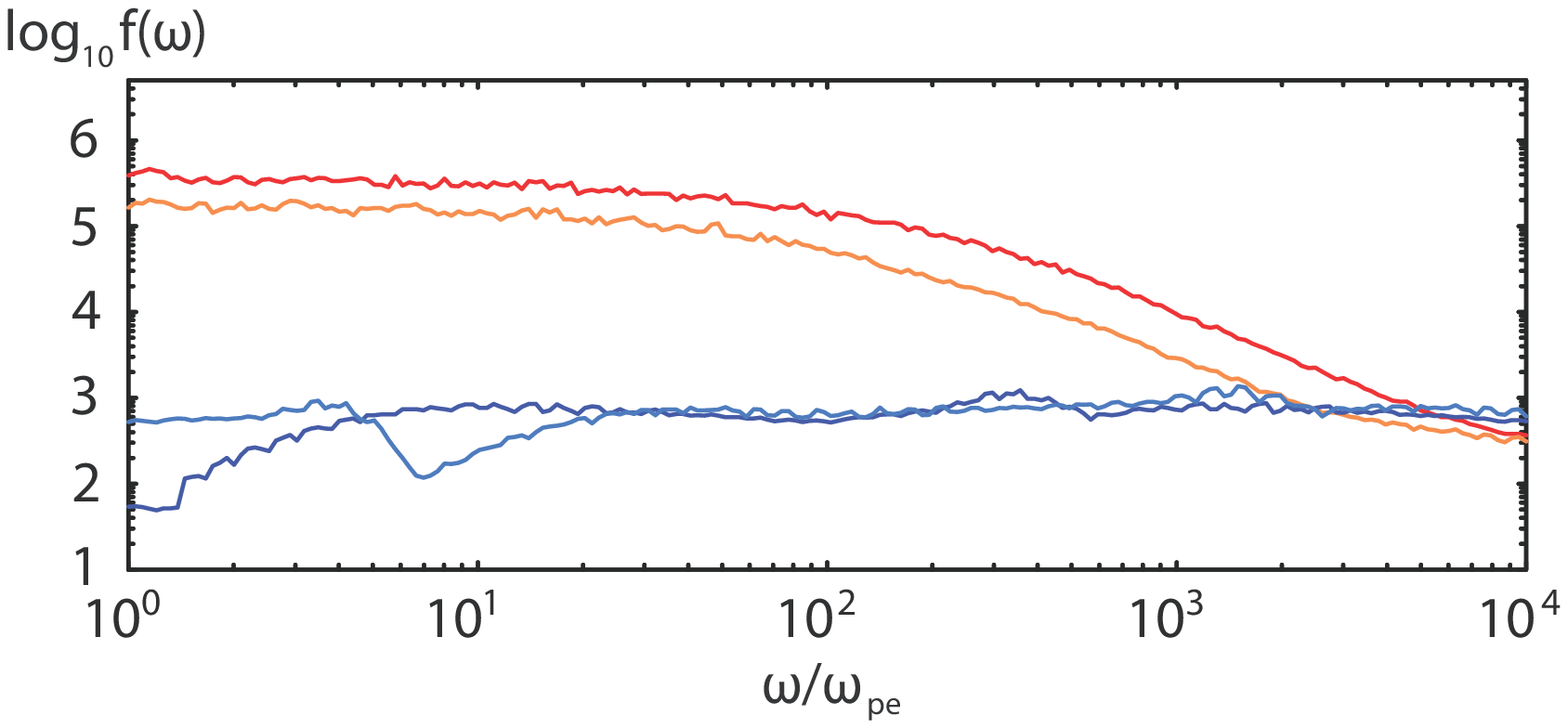}
\includegraphics[width=80mm]{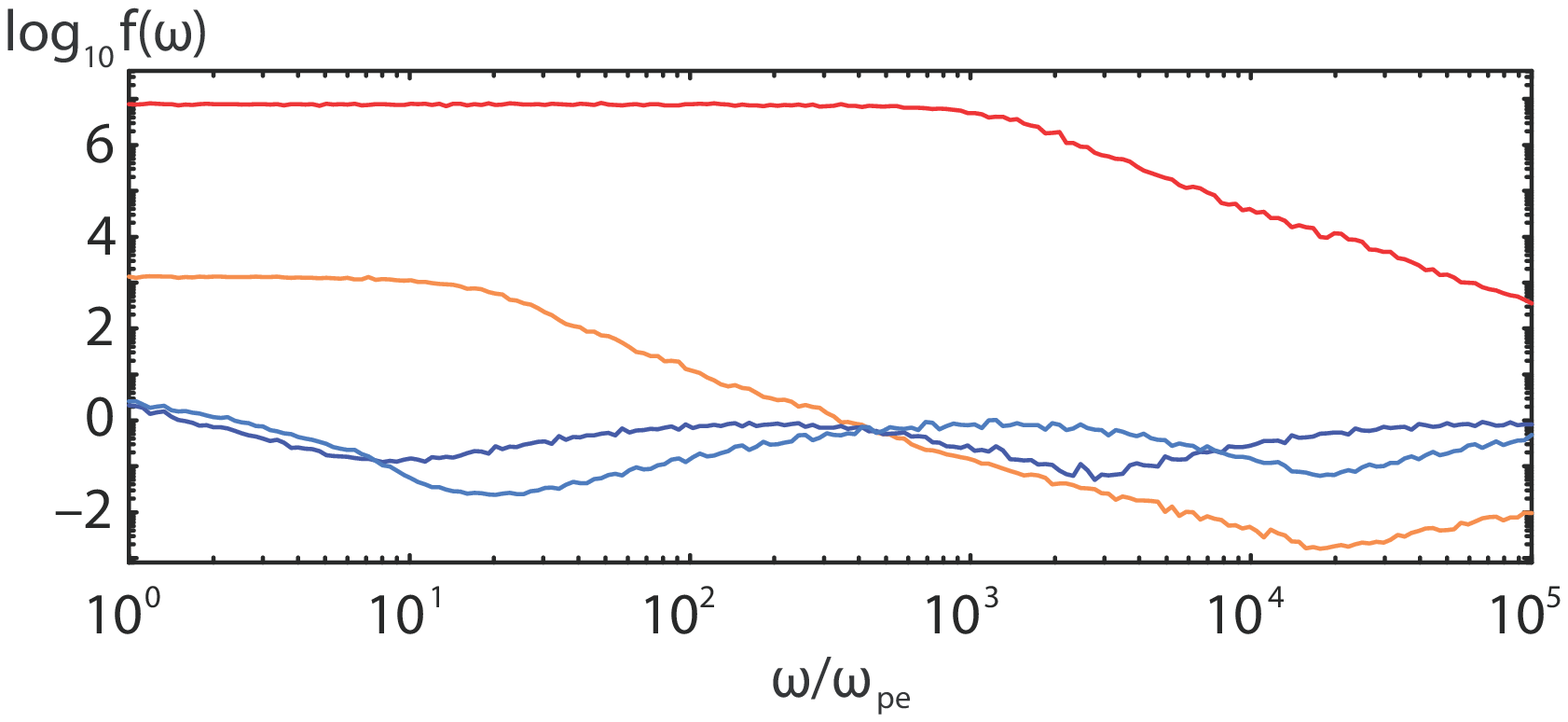}

\vspace{-0.cm}
\caption{Spectra obtained from jet and ambient electrons for the two viewing angles for $\gamma = 15$ (upper)
and $\gamma = 100$ (lower).  
Spectra with jet electrons are shown in red ($0^{\circ}$) and orange ($5^{\circ}$). 
Spectra from ambient electrons show the lowest levels by blue ($0^{\circ}$) and light blue 
($5^{\circ}$).}
\end{figure}

The spectra shown in Fig. 5 are obtained for emission from jet electrons and ambient electrons 
separately for two case with $\gamma = 15 {\rm and} 100$. In this case the spectra are calculated for head-on radiation ($0^{\circ}$) and $5^{\circ}$. It is noted that In the case with $\gamma = 100$ spectrum is extended
in the higher frequency. However, the spectrum with viewing angle $5^{\circ}$ is greatly decreased in particular
in the high frequency due to the narrow beaming angle. 

The radiation from jet electrons show Bremsstrahlung-like spectra as a red line 
($0^{\circ}$) and orange line ($5^{\circ}$) ~\cite{hedeT05}. The spectra with jet electrons are 
different from the spectra shown in Fig. 2c.
Since the magnetic fields generated by the Weibel instability are rather weak and the jet electrons 
are not much accelerated, the trajectories of jet electrons are almost straight with only a slight 
bent.  

\begin{figure}[!h]
\centering
\includegraphics[width=60mm]{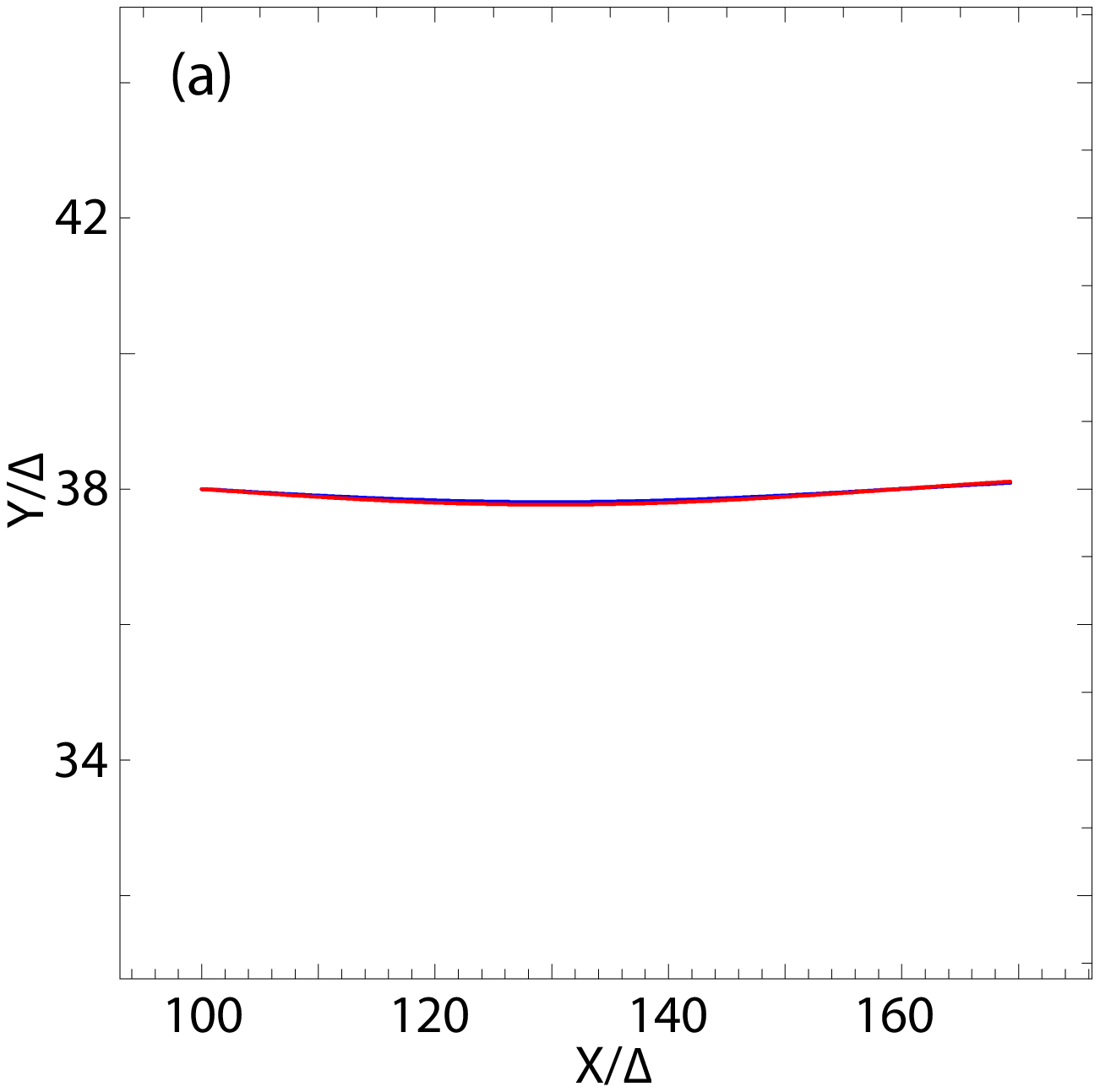}
\includegraphics[width=65mm]{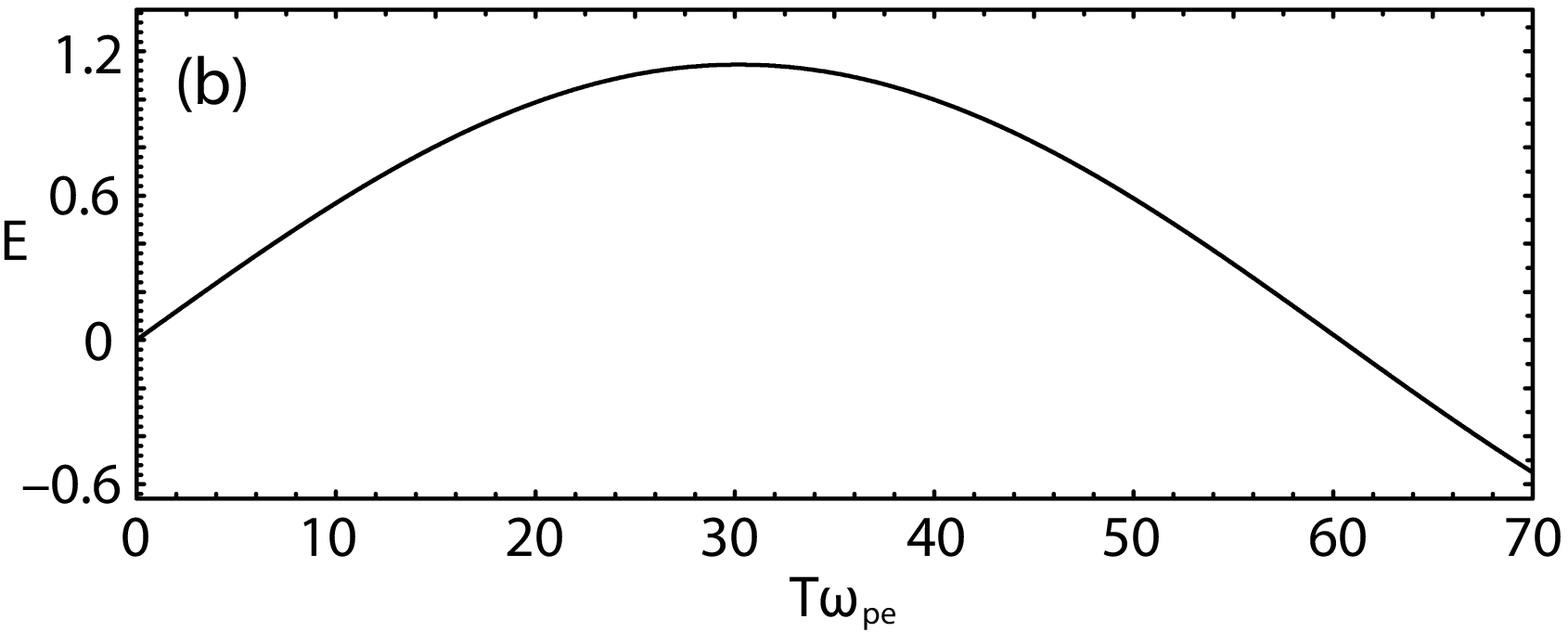}
\includegraphics[width=75mm]{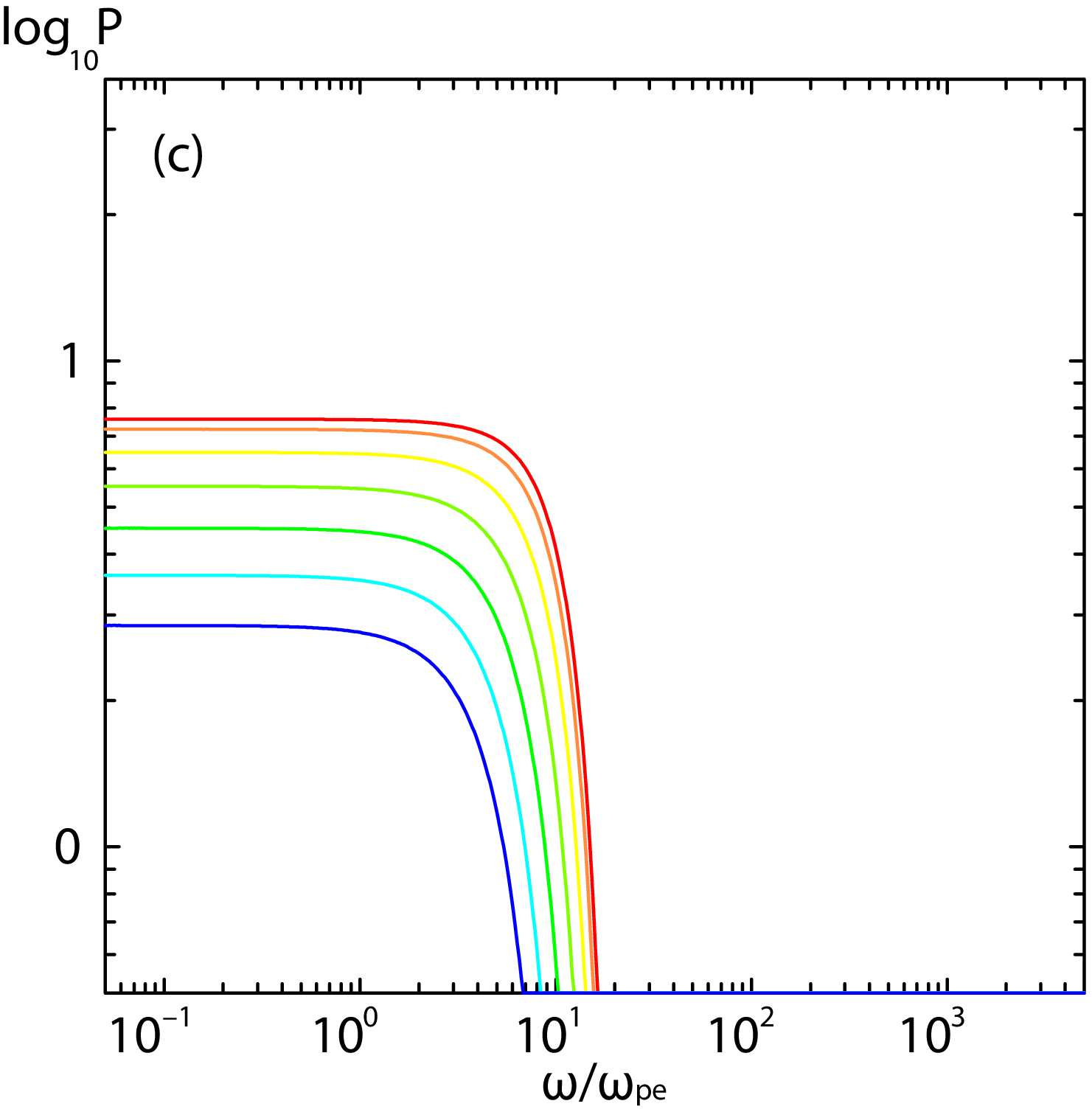}
\vspace*{-0.cm}
\caption{The case with a weak magnetic field ($B_{\rm x} = 0.37$) and small 
perpendicular velocity  ($v_{\perp 1} = 0.01c, v_{\perp 2} = 0.012c$).
The paths of two electrons moving helically along the $x-$direction in a homogenous magnetic
field shown in the $x-y$-plane (a). The two electrons radiate a time dependent electric field. 
An observer situated at great distance along the n-vector sees the retarded electric field from
the moving electrons (b). 
The observed power spectrum at different viewing angles from the two electrons (c). 
Frequency is in units of $\omega_{\rm pe}^{-1}$. }
\end{figure}

\vspace{-0.cm}
We compare these spectra with our known spectra obtained from two (jet) electrons, the case 
with a parallel magnetic field ($B_{\rm x} =0.37$), and jet velocity of $v_{\rm j1,2} = 0.99c$. 
Two electrons 
are injected with different perpendicular velocities ($v_{\perp 1} = 0.01c, v_{\perp 2} 
= 0.012c$). A maximum Lorenz factor of 
$\gamma_{\max} =\{(1 - (v_{\rm j2}^{2} +v_{\perp 2}^{2})/c^{2}\}^{-1/2}  
= 7.114$ accompanies the larger perpendicular velocity. The critical angle for off-axis
radiation $\theta_{\gamma} = 180^{\circ}/(\pi \gamma_{\max})$ for this case is
8.05$^{\circ}$. 

Comparing the spectra with Figs. 5 and 6c we find similarities. The lower frequencies have flat 
spectra and the higher frequencies decrease monotonically. The slope in Fig. 5 is less steep than 
that in Fig. 6c. This is due to the fact that the spread of Lorenz factors of jet electrons is 
substantial  and the average Lorenz factor is larger as well. 
As shown in Fig. 7.16  in Hededal's
Ph. D. thesis ~\cite{hedeT05},  the turbulent magnetic field with the red noise ($\mu = -3$) makes 
the spectrum shifted toward higher frequencies. This effect is found in Fig. 5

 We obtained spectra using several different parameters with 
jet electrons and ambient magnetic field. However, the strength of the magnetic 
fields generated by the Weibel instability is small  in the region $x/\Delta < 500$ 
($\epsilon_{\rm B} < 0.07$) as shown in Fig. 1b, therefore the spectra for 
these cases are very similar to Fig. 5. As shown in Fig. 7.12 in Hededal's Ph. D. 
thesis ~\cite{hedeT05}, the trajectories of jet electrons have to be chaotic to 
produce a jitter-like spectrum  as shown in Fig. 7.22. 

In order to obtain the spectrum of synchrotron (jitter) emission, we 
consider an ensemble of electrons selected in the region where the
Weibel instability has fully grown and electrons are accelerated in
the generated magnetic fields as shown in Fig. 1, which is being investigted. 

\section{DISCUSSIONS}
Emission obtained with the method described above is obtained self-consistently,
and automatically accounts for magnetic field structures on small scales responsible 
for jitter emission. By performing such calculations for simulations with different 
parameters, we can investigate and compare the different regimes of jitter- and
synchrotron-type emission ~\cite{medv00,medv06}. 
The feasibility of this approach has already been demonstrated ~\cite{hedeT05,hedeN05}, 
and its implementation is straightforward. Thus, we should be able to address the low 
frequency GRB spectral index violation of the synchrotron spectrum line of death ~\cite{medv06}. 

Medvedev and Spitkovsky recently showed that electrons may cool efficiently at or near the shock 
jump and are capable of emitting a large fraction of the shock energy ~\cite{MS09}. Such shocks 
are well-resolved in existing PIC simulations; therefore, the microscopic structure can be studied
in detail. Since most of the emission in such shocks would originate from the
vicinity of the shock, the spectral power of the emitted radiation can be directly
obtained from finite-length simulations and compared with observational data.

As shown in Fig. 1, behind the trailing shock the electrons are accelerated and strong magnetic 
fields are generated. Therefore, this region seems to produce the emission that is observed by 
satellites. We will calculate more spectra based on our RPIC simulations and compare in detail 
with Fermi data.

\bigskip 
\begin{acknowledgments}
This work is supported by NSF-AST-0506719, 
AST-0506666, AST-0908040, AST-0908010, NASA-NNG05GK73G, NNX07AJ88G, NNX08AG83G, NNX08AL39G, and NNX09AD16G.
 JN was supported by MNi-SW research 
projects 1 P03D 003 29 and N N203 393034,
 and The Foundation for Polish Science through the HOMING program, which is
 supported through the EEA Financial Mechanism. Simulations were performed at the  
Columbia facility at the NASA
Advanced Supercomputing (NAS).  and SGI Altix (obalt) at the National
Center for Supercomputing Applications (NCSA) which is supported by
the NSF. 
Part of this work was done while K.-I. N. was visiting the
Niels Bohr Institute. Support from the Danish Natural Science Research Council is gratefully acknowledged.
This report was finalized during the program ``Particle Acceleration in Astrophysical Plasmas'' at the Kavli Institute 
for Theoretical Physics which is supported by  the National Science Foundation under Grant No. PHY05-51164.

\end{acknowledgments}

\bigskip 

\end{document}